\documentclass[a4paper,11pt]{article}
\usepackage{graphicx}
\usepackage{latexsym}
\usepackage{amsmath,amsthm}
\usepackage{amssymb}

\title{\bf An application of the Casoratian technique \\ to the 2D Toda lattice equation}
\date{}

\author{
Wen-Xiu Ma$^{a,b}$ \thanks{Email: {\tt mawx@cas.usf.edu}}
\\
{\small $^a$Department of Mathematics and Statistics, University of South Florida,}\\
{ \small Tampa, FL 33620-5700, USA}
\\{ \small
$^b$State Key
   Laboratory of Scientific and Engineering Computing,}
  \\
  {\small  Chinese Academy of
   Sciences,
   P.O. Box 2719, Beijing 100080, PR China
}
}

\begin{document}

\maketitle


\setlength{\baselineskip}{17.5pt}


\begin{abstract}
A general Casoratian formulation is proposed for the 2D Toda lattice
equation, which involves coupled eigenfunction systems. Various
Casoratian type solutions are generated, through solving the
resulting linear conditions and using a B\"acklund transformation.


\end{abstract}


\vskip 0.5cm

\def \be {\begin{equation}}
\def \ee {\end{equation}}
\def \bea {\begin{eqnarray}}
\def \eea {\end{eqnarray}}
\def \ba {\begin{array}}
\def \ea {\end{array}}
\def \D {\displaystyle }

\def \part {\partial }

\newcommand{\R}{\mathbb{R}}
\newcommand{\Z }{\mathbb{Z}}

\newtheorem{theorem}{Theorem}
\newtheorem{lemma}{Lemma}
\newtheorem{col}{Corollary}

\def\cdot{{\scriptstyle\,\bullet\,}}

\newpage

\section{\bf Introduction}

It is well-known that Wronskian formulations show a common
characteristic feature of continuous soliton equations, and provide
a powerful tool to construct exact solutions to continuous soliton
equations \cite{FN}-\cite{LiMLZ-IP2007}. The resulting technique
has
applied to many continuous soliton equations such as the KdV, MKdV,
NLS, derivative NLS, Boussinesq, KP, sine-Gordon and sinh-Gordon
equations. With Wronskian formulations, soliton solutions and
rational solutions are usually expressed as some kind of logarithmic
derivatives of Wronskian type determinants with respect to space
variables, and the involved determinants are generated by
eigenfunctions satisfying linear systems of differential equations.
A great help is that Wronskian formulations transform nonlinear
problems into linear problems, and thus continuous soliton equations
can be treated by means of linear theories.

There is a discrete version of Wronskian formulations, called
Casoratian formulations, for discrete soliton equations such as the
Volttera, nonlinear electrical network, and Toda lattice equations
(see, for example, \cite{MaM-PA2004}-\cite{WangHG-JCAM2007}). With
Casoratian formulations, soliton solutions and rational solutions
are often expressed as some kind of rational functions of Casoratian
type determinants, and the involved determinants are made of
eigenfunctions satisfying linear systems of differential-difference
equations. Therefore, the Casoratian technique offers a direct
approach for constructing exact solutions to discrete soliton
equations.

Besides soliton solutions and rational solutions, the Wronskian and
Casoratian techniques can be used to construct positon solutions
\cite{ArkadievPP-ZNSL1984}-\cite{MarunoMO-JPSJ2004}, i.e., solutions
involving one kind of transcendental functions: trigonometric
functions. More generally, a novel kind of solutions called
compelxiton solutions has been introduced and generated using such
techniques for continuous and discrete soliton equations
\cite{Ma-PLA2002,MaM-PA2004} and soliton equations with sources
\cite{Ma-CSF2005}. Those solutions contain two kinds of
transcendental waves: exponential waves and trigonometric waves,
with different speeds, and they correspond to complex eigenvalues of
associated characteristic linear problems and generate solitons and
positons as limit cases of the complex eigenvalues
\cite{MaY-TAMS2005,Ma-NA2005}.

One of intriguing discrete soliton equations is the 2D Toda lattice
equation \cite{Hirota-book2004}
\begin{equation}  \frac {\part ^2 Q_n}{\part s \part x}
=V_{n+1}-2V_n+V_{n-1},\ Q_n=\ln (1+V_n),
\label{eq:2dToda:pma-2dToda}
\end{equation}
where $x,s\in \R $ and $n\in \Z $. Through the dependent variable
transformation \be V_n=\frac {\part ^2 }{\part s\part x} \ln \tau
_n,  \ee  the equation \eqref{eq:2dToda:pma-2dToda} may be
integrated with respect to $x$ and $s$ to obtain \be 1+ \frac {\part
^2 }{\part s\part x}\ln \tau _n =\frac {\tau_{n+1} \tau _{n-1}}{\tau
_n^2}, \label{eq:db2dToda:pma-2dToda} \ee where the constants of
integration are set to zero. This equation is equivalent to \be
\frac {\part ^2 \tau _n}{\part s\part x}\tau _n -\frac {\part \tau
_n}{\part s}\frac {\part \tau _n}{\part x}=\tau _{n+1}\tau
_{n-1}-\tau _n^2, \label{eq:b2dToda:pma-2dToda} \ee which can be
written as \be D_xD_s \tau _n\cdot \tau_n =2(\tau _{n+1}\tau
_{n-1}-\tau _n^2), \label{eq:Hirotabilinearform2dToda:pma-2dToda}
\ee in terms of Hirota's operator \cite{Hirota-book2004}: \be
(D_zf\cdot g )=(\part _z-\part _{z'})f(z)g(z')|_{z'=z}. \ee If we
set \be y_n=\ln \frac {\tau _n}{\tau _{n+1}} ,
\label{eq:Transformation2of2dToda:pma-2dToda}
 \ee then we obtain another
form for the 2D Toda lattice equation: \be \frac {\part ^2 y
_n}{\part s\part x} = {\rm e} ^{y_{n-1}-y_n}- {\rm e
}^{y_n-y_{n+1}}. \label{eq:2ndformof2dToda:pma-2dToda} \ee Two forms
\eqref{eq:2dToda:pma-2dToda} and
\eqref{eq:2ndformof2dToda:pma-2dToda} of the 2D Toda lattice
equation are linked through
\[ \frac {\part ^2 y
_n}{\part s\part x} =V_n-V_{n+1}.\]
 In this paper, we would like to establish a general Casoratian
formulation for the 2D Toda lattice equation
\eqref{eq:b2dToda:pma-2dToda} and analyze its exact solutions based
on the resulting Casoratian formulation and a B\"acklund
transformation.

The paper is organized as follows. In Section
\ref{sec:CasoratianFormulation:pma-2dToda}, a general Casoratian
formulation is presented for the bilinear 2D Toda lattice equation
\eqref{eq:b2dToda:pma-2dToda}. In Section
\ref{sec:CasoratianSolutions:pma-2dToda}, some specific cases of
linear conditions are discussed and a B\"acklund transformation is
furnished to construct exact solutions, and various examples of
Casoratian type solutions are presented. Concluding remarks are
given finally in Section \ref{sec:Concludingremarks:pma-2dToda}.

\section{A general Casoratian formulation}
\label{sec:CasoratianFormulation:pma-2dToda}

The $N$-soliton solution to the bilinear 2D Toda lattice equation
\eqref{eq:b2dToda:pma-2dToda} is expressed as a Casorati determinant
\cite{HirotaOS-JPSJ1988} \be \tau
_n=\textrm{Cas}(\phi_1,\phi_2,\cdots,\phi_N)
=\left|  \ba {cccc} \phi_1(n) &\phi_1(n+1) & \cdots & \phi_1(n+N-1) \vspace{2mm}\\
\phi_2(n) &\phi_2(n+1) & \cdots & \phi_2(n+N-1) \vspace{2mm}\\
\vdots &\vdots &  & \vdots \vspace{2mm}\\
\phi_N(n) &\phi_N(n+1) & \cdots & \phi_N(n+N-1)
 \ea \right|  ,  \label{eq:CasDet:pma-2dToda}  \ee
where each $\phi_i(n)=\phi_i(n,x,s)$ satisfies the linear
differential-difference equations \be  \frac {\part \phi_i(n)}{\part
x} =\phi_i(n+1),\ \frac {\part \phi_i(n)}{\part s} =-\phi_i(n-1),\
1\le i\le N. \label{eq:linearconditions:pma-2dToda} \ee The above
Casorati determinant has been used in the theory of the 1D lattice
equations \cite{MaM-PA2004}-\cite{WangHG-JCAM2007}. We will adopt
the notation \cite{MaM-PA2004} \be k..l=k,k+1,\cdots,l
 \ee where $ k< l$, and denote
 the generalized Casorati determinant by
\be |i_1,\cdots,i_N| = \det ([i_1,\cdots,i_N]), \ee where $ i_j\in
\Z ,\ 1\le j\le  N$, and the matrix $[i_1,\cdots,i_N]$
is defined by
 \be \ba {l}
[i_1,\cdots,i_N]  =\left[  \ba {cccc} \phi_1(n+i_1) & \phi_1(n+i_2)
& \cdots & \phi_1(n+i_N)
 \vspace{2mm}\\
\phi_2(n+i_1) &\phi_2(n+i_2) & \cdots & \phi_2(n+i_N)  \vspace{2mm}\\
\vdots &  \vdots &  &\vdots \vspace{2mm}\\
\phi_N(n+i_1) & \phi_N(n+i_2)&\cdots &\phi_N(n+i_N) \ea \right]. \ea
\ee  Obviously, the standard Casorati determinant is given by
\[ \textrm{Cas}(\phi_1,\phi_2,\cdots,\phi_N)=|0..N-1|.\]

\begin{theorem}\label{thm:gsoltau:pma-2dToda}
Let $\varepsilon =\pm 1$ and $\delta =\pm 1$, i.e., $(\varepsilon
,\delta )=(1,1),\ (1,-1),\ (-1,1) $ or $ (-1,-1)$. If a set of
functions $\phi _i(n)=\phi_i(n,x,s),\ 1\le i\le N,$ satisfies the
following coupled linear differential-difference equations: \bea &&
\frac {\part \phi_i(n)}{\part x} = \varepsilon \phi_i(n+\delta )
+\sum_{j=1}^N\lambda _{ij}(x)\phi _j(n) ,
\ 1\le i\le N,\label{eq:cond1:pma-2dToda} \\
&& \frac {\part \phi_i(n)}{\part s} =- \varepsilon  \phi_i(n-\delta
) + \sum_{j=1}^N\mu _{ij}(s)\phi _j(n) ,\ 1\le i\le
N,\label{eq:cond2:pma-2dToda} \eea where  $\lambda _{ij}(x)$ and
$\mu _{ij}(s)$, $1\le i,j\le N$, are arbitrary real functions, then
$\tau _n= |0..N-1|$ defined by \eqref{eq:CasDet:pma-2dToda} solves
the bilinear 2D Toda lattice equation \eqref{eq:b2dToda:pma-2dToda}.
\end{theorem}

\noindent {\it Proof:} Under an exchange of the variables $x$ and
$s$, the cases of linear conditions \eqref{eq:cond1:pma-2dToda} and
\eqref{eq:cond2:pma-2dToda} with different values $\delta=\pm 1 $
are transformed into each other, but the bilinear 2D Toda lattice
equation \eqref{eq:b2dToda:pma-2dToda} is invariant. Therefore, we
only need to check the case under $\delta =1$. In what follows, we
set $\delta =1$.

Let us use $(Ef)(n)=f(n+1)$ and define \be
(L_x\phi_i)(n)=\sum_{i=1}^N\lambda _{ij}\phi_j(n), \
(L_s\phi_i)(n)=\sum_{i=1}^N\mu _{ij}\phi_j(n), \ 1\le i\le N.  \ee
Then, using \eqref{eq:cond1:pma-2dToda}, we can compute that
\[ \ba {l}\D
\frac {\part \tau _n}{\part x}
=\sum_{i=1}^N \left|  \ba {cccc} \phi_1(n) &\phi_1(n+1) & \cdots & \phi_1(n+N-1) \vspace{2mm}\\
\vdots &\vdots &  & \vdots \vspace{2mm}\\
\D \part _x \phi_i (n) & \part _x \phi_i (n+1)  & \cdots  & \part _x \phi_i (n+N-1) \vspace{2mm}\\
\vdots &\vdots &  & \vdots \vspace{2mm}\\
\phi_N(n) &\phi_N(n+1) & \cdots & \phi_N(n+N-1)
 \ea \right|
\vspace{2mm} \\
\D\quad  = \varepsilon
\sum_{i=1}^N \left|  \ba {cccc} \phi_1(n) &\phi_1(n+1) & \cdots & \phi_1(n+N-1) \vspace{2mm}\\
\vdots &\vdots &  & \vdots \vspace{2mm}\\
 (E\phi_i) (n) & (E\phi_i) (n+1) &  \cdots & (E\phi_i)(n+N-1) \vspace{2mm}\\
\vdots &\vdots &  & \vdots \vspace{2mm}\\
\phi_N(n) &\phi_N(n+1) & \cdots & \phi_N(n+N-1)
 \ea \right|
\vspace{2mm}\\
\D\quad\quad   +\sum_{i=1}^N \left|  \ba {cccc} \phi_1(n) &\phi_1(n+1) & \cdots & \phi_1(n+N-1) \vspace{2mm}\\
\vdots &\vdots &  & \vdots \vspace{2mm}\\
 (L_x\phi_i)(n) & (L_x\phi_i)(n+1)  & \cdots &  (L_x \phi_i) (n+N-1) \vspace{2mm}\\
\vdots &\vdots &  & \vdots \vspace{2mm}\\
\phi_N(n) &\phi_N(n+1) & \cdots & \phi_N(n+N-1)
 \ea \right|
 \vspace{2mm}\\
\D \quad = \varepsilon \sum_{j=1}^N \left|  \ba {cccccc} \phi_1(n) &
\phi_1(n+1) & \cdots &
(E\phi_1)(n+j-1)&  \cdots & \phi_1(n+N-1) \vspace{2mm}\\
\phi_2(n) & \phi_2(n+1) & \cdots &
(E\phi_2)(n+j-1)&  \cdots & \phi_2(n+N-1) \vspace{2mm}\\
\vdots &\vdots &  & \vdots & & \vdots\vspace{2mm}\\
\phi_N(n) &\phi_N(n+1) & \cdots & (E\phi_N)(n+j-1) & \cdots &
\phi_N(n+N-1)
 \ea \right| \vspace{2mm}\\
\D \quad \quad +\sum_{i=1}^N \left|  \ba {cccc} \phi_1(n)
&\phi_1(n+1)
& \cdots & \phi_1(n+N-1) \vspace{2mm}\\
\vdots &\vdots &  & \vdots \vspace{2mm}\\
 \lambda _{ii} \phi_i(n) & \lambda _{ii} \phi_i(n+1)  & \cdots  & \lambda _{ii}  \phi_i (n+N-1) \vspace{2mm}\\
\vdots &\vdots &  & \vdots \vspace{2mm}\\
\phi_N(n) &\phi_N(n+1) & \cdots & \phi_N(n+N-1)
 \ea \right|
\vspace{2mm}\\
\D \quad = \varepsilon |0..N-2,N|+ \bigl(\sum_{i=1}^N\lambda
_{ii}\bigr)\tau _n.
 \ea \]
Using almost the same argument, we can obtain \[  \frac {\part \tau
_n}{\part s} =-\varepsilon |-1,1..N-1|+\bigl(\sum_{i=1}^N\mu
_{ii}\bigr)\tau_n. \] Further, we can similarly compute that \[ \ba
{l}\D \frac {\part ^2 \tau _n}{\part s\part x} =-|-1,1..N-2,N|-\tau
_n+ \varepsilon \bigl(\sum_{i=1}^N\mu _{ii}\bigr) |0..N-2,N|
\vspace{2mm}\\
\D\quad  +\bigl(\sum_{i=1}^N\lambda _{ii}\bigr)\bigl[ - \varepsilon
|-1,1..N-1|+\bigl(\sum_{i=1}^N\mu _{ii} \bigr)\tau_n\bigr]
\vspace{2mm}\\
\D =-|-1,1..N-2,N| -\tau _n + \varepsilon \bigl(\sum_{i=1}^N\mu
_{ii}\bigr) |0..N-2,N|
 \vspace{2mm}\\
\quad \D  - \varepsilon \bigl(\sum_{i=1}^N\lambda _{ii}\bigr)
|-1,1..N-1|+ \bigl(\sum_{i=1}^N\lambda _{ii}\bigr)
\bigl(\sum_{i=1}^N\mu _{ii}\bigr) \tau _n.
 \ea \]
Plugging these results into the bilinear equation
\eqref{eq:b2dToda:pma-2dToda} gives \[ \ba {l}\D  \frac {\part ^2
\tau _n}{\part s\part x}\tau _n -\frac {\part \tau _n}{\part s}\frac
{\part \tau _n}{\part x}-\tau _{n+1}\tau
_{n-1}+\tau _n^2\vspace{2mm}\\
\D = -|-1,1..N-2,N||0..N-1|+|0..N-2,N||-1,1..N-1|-|1..N||-1..N-2|.
\ea \] This sum is the Laplace expansion by $N\times N$ minors of
the following $2N\times 2N$ determinant:
\[ -\frac 12 \left| \ba {rcclrccl} [ \hspace{-2mm} & -1,0 , &  1..N-2 & \hspace{-2mm} \left. \right ]
& \left [\right. \hspace{-2mm}
 & \emptyset, &  N-1, N & \hspace{-2mm} \left. \right ] \vspace{2mm}\\
\left[ \right. \hspace{-2mm} & -1, 0, &  \emptyset  & \hspace{-2mm}
\left. \right] & [ \hspace{-2mm} & 1..N-2, & N-1, N & \hspace{-2mm}
]
  \ea \right|
,
\]
where $\emptyset$ indicates the $N\times (N-2)$ zero matrix, and
$[\emptyset, N-1,N]=[\emptyset, \Phi(n+N-1),\Phi(n+N)]$ and $[-1,0,
\emptyset]=[\Phi(n-1),\Phi(n),\emptyset  ] $ with
$\Phi(m)=(\phi_1(m),\cdots,\phi_N(m))^T$. Obviously, this
determinant is zero. Therefore, the solution is verified. \hfill
$\Box$

The linear conditions \eqref{eq:cond1:pma-2dToda} and
\eqref{eq:cond2:pma-2dToda} in the case of
$(\varepsilon,\delta)=(1,1)$ is a generalization of the conditions
\eqref{eq:linearconditions:pma-2dToda}. Theorem
\ref{thm:gsoltau:pma-2dToda} tells us that if a set of functions
$\phi_i(n), \ 1\le i\le N,$ satisfies all linear conditions in
\eqref{eq:cond1:pma-2dToda} and \eqref{eq:cond2:pma-2dToda}, then we
can get a Casoratian solution $\tau _n=|0..N-1|$ to the bilinear 2D
Toda lattice equation \eqref{eq:b2dToda:pma-2dToda}. If we exchange
$x$ and $s$ in $\tau _n$, we can get another Casoratian solution,
based on Theorem \ref{thm:gsoltau:pma-2dToda}.

Let us observe how the Casoratian formulation generates solutions a
little bit more carefully. From the compatibility conditions
$\phi_{i,xs}=\phi_{i,sx}$, $1\le i\le N$, of the conditions
\eqref{eq:cond1:pma-2dToda} and \eqref{eq:cond2:pma-2dToda}, we have
the equalities \be \sum_{j,k=1}^N(\lambda_{ij}\mu_{jk}-\mu_{ij}
\lambda_{jk} )\phi_k=0, \  1\le i\le N, \ee
and thus we see that the
Casorati determinant $\mbox{Cas}(\phi_1,\phi_2,\cdots,\phi_N)$
becomes zero at a point $(x,s)$ where the coefficient matrices
$A=A(x)=(\lambda_{ij}(x))_{N\times N}$ and $B=B(s) =(\mu
_{ij}(s))_{N\times N}$ do not commute. Therefore, if $A$ and $B$ are
constant and don't commute, then $\tau _n=|0..N-1|$ is zero. This
shows that
 the reduced case of \eqref{eq:cond1:pma-2dToda} and
\eqref{eq:cond2:pma-2dToda} under
\begin{equation}
A(x) B(s)-B(s)A(x) =0
\label{eq:condition1forspecialsolutionofphi_i:pma-2dToda}
\end{equation} is important in generating non-trivial Casoratian
solutions to the bilinear 2D Toda lattice equation
\eqref{eq:b2dToda:pma-2dToda}.

\section{\bf Casoratian type solutions}
\label{sec:CasoratianSolutions:pma-2dToda}

We would like to construct exact solutions of the bilinear 2D Toda
lattice equation \eqref{eq:b2dToda:pma-2dToda} by using the
resulting Casoratian formulation and introducing a B\"acklund
transformation.

\begin{theorem}\label{thm:CasoratianSolutions:pma-2dToda}
If $A(x)=(\lambda _{ij}(x))_{N\times N}$ and $B(s)=(\mu
_{ij}(s))_{N\times N}$ are continuous and satisfy
\eqref{eq:condition1forspecialsolutionofphi_i:pma-2dToda} and
\begin{eqnarray}
 && A(x) \int_0^x A(x')\, dx'=\int_0^x A(x')\, dx' \, A(x),
 \label{eq:condition2forspecialsolutionofphi_i:pma-2dToda} \\
 && B(s) \int_0^s B(s')\, ds'= \int_0^s B(s')\, ds' \, B(s) ,
\label{eq:condition3forspecialsolutionofphi_i:pma-2dToda}
 \end{eqnarray} then
the linear differential-difference equations
\eqref{eq:cond1:pma-2dToda} and \eqref{eq:cond2:pma-2dToda} have the
following solution \be \Phi=\Phi(n)= \exp  ( \int_0^x A(x')\, dx' +
\int_0^s B(s')\, ds')(p_1^n\,{\rm e}^{\, \varepsilon (p_1^\delta x
-p_1^{-\delta }s)+q_1},\cdots , p_N^n\,{\rm e}^{\,\varepsilon
(p_N^\delta x-p_N^{-\delta }s)+q_N} )^T,
\label{eq:ssolutionofphi_i:pma-2dToda} \ee
 where $\Phi=(\phi_1,\cdots,\phi_N)^T$ and
$p_i\ne 0,\ q_i,\ 1\le i\le N$, are arbitrary real constants.
 \end{theorem}

\noindent {\it Proof:} The condition
\eqref{eq:condition1forspecialsolutionofphi_i:pma-2dToda} implies
that
\[\ba {l} \D \quad  \exp ( \int_0^x A(x')\, dx' + \int_0^s B(s')\,
ds')\vspace{2mm}\\
\D  =  \exp  (\int_0^x A(x')\, dx' ) \exp  ( \int_0^s B(s')\, ds') \vspace{2mm}\\
\D
= \exp ( \int_0^s B(s')\, ds') \exp  (\int_0^x A(x')\, dx' )
,\vspace{2mm}\\
\D  A(x) \exp  (\int_0^s B(s')\, ds' )=\exp  (\int_0^s B(s')\, ds'
)A(x),\vspace{2mm}\\
\D  B(s) \exp  (\int_0^x A(x')\, dx' )=\exp (\int_0^x A(x')\, dx'
)B(s). \ea \] The other two conditions
\eqref{eq:condition2forspecialsolutionofphi_i:pma-2dToda} and
\eqref{eq:condition3forspecialsolutionofphi_i:pma-2dToda} guarantee
that
\begin{eqnarray} &&
\partial _x \exp  (\int_0^x A(x')\, dx' ) =A(x) \exp  (\int_0^x A(x')\, dx'
), \label{eq:fundamentalsolutiononx:pma-2dToda} \\
&&  \partial _x \exp  (\int_0^s B(s')\, ds' ) =B(s) \exp (\int_0^s
B(s')\, ds'), \label{eq:fundamentalsolutionons:pma-2dToda}
\end{eqnarray}
respectively. Further, a direct computation shows that
\[
\frac {\part  \Phi(n) }{\part x} = \varepsilon \Phi(n+\delta ) +
A(x) \Phi(n)  , \ \frac {\part  \Phi(n) }{\part s} = - \varepsilon
\Phi(n-\delta ) + B(s) \Phi(n) .
 \]
This verifies the solution in
\eqref{eq:ssolutionofphi_i:pma-2dToda}. \hfill $\Box$

Noting that \eqref{eq:cond1:pma-2dToda} and
\eqref{eq:cond2:pma-2dToda} are linear, any linear combination of
$\Phi$ defined by \eqref{eq:ssolutionofphi_i:pma-2dToda} with
different sets of $p_i$ and $q_i$, $1\le i\le N$, is again a
solution to \eqref{eq:cond1:pma-2dToda} and
\eqref{eq:cond2:pma-2dToda}. One example is the set of functions \be
\phi_i=\sum_{j=1}^M p_{ij}^n\textrm{e}^{\,(\varepsilon p_{ij}^\delta
+\lambda _i) x-(\varepsilon p_{ij}^{-\delta }-\mu _i)s+q _{ij}}, \
1\le i\le N, \label{eq:specialcase1ofphi_i:pma-2dToda}  \ee where
the $p_{ij}$'s are arbitrary non-zero real constants and the
$q_{ij}$'s, $\lambda_i$'s and $\mu_{i}$'s are arbitrary real
constants. Actually, $\Phi=(\phi_1,\cdots, \phi_N)^T$ satisfies the
linear conditions \eqref{eq:cond1:pma-2dToda} and
\eqref{eq:cond2:pma-2dToda} with $A=\textrm{diag}(\lambda
_1,\cdots,\lambda _N)$ and $B=\textrm{diag}(\mu _1,\cdots,\mu _N)$.
Thus, we have a Casoratian solution
$\tau_n=|0..N-1|=\textrm{Cas}(\phi_1,\cdots,\phi_N)$. The
$N$-soliton solutions correspond to $M=2$ \cite{Hirota-PTPS1988}.
The situation with a general integer $M$ yields new Casoratian
solutions involving many free parameters.

If for each $l\le N$, we further take $\lambda _i=\mu _i,\ 1\le i\le
l$, then \be \Phi= (\phi_1,
\part _{\lambda _1} \phi_1 ,\cdots, \frac 1 {k!}\part _{\lambda
_1}^{k_1} \phi_1 ;\cdots; \phi_l,
\part _{\lambda _l} \phi_l ,\cdots, \frac 1 {k!}\part _{\lambda
_l}^{k_l} \phi_l ), \label{eq:specialcase2ofphi_i:pma-2dToda} \ee
where $k_1+\cdots +k_l=N$, satisfies the linear conditions
\eqref{eq:cond1:pma-2dToda} and \eqref{eq:cond2:pma-2dToda} with
\[ A=B=\textrm{diag}(C_1,\cdots ,C_l ),\ C_i=\left[\ba {cccc}   \lambda _i & & & 0 \vspace{2mm}\\
1& \lambda _i & & \vspace{2mm}\\
& \ddots &\ddots & \vspace{2mm}\\
0& & 1 & \lambda _i
   \ea \right] ,\ 1\le i\le l.  \]
Thus, this gives us the following Casoratian solution \be \tau _n=
\textrm{Cas}(\phi_1,
\part _{\lambda _1} \phi_1 ,\cdots, \frac 1 {k!}\part _{\lambda
_1}^{k_1} \phi_1 ;\cdots; \phi_l,
\part _{\lambda _l} \phi_l ,\cdots, \frac 1 {k!}\part _{\lambda
_l}^{k_l} \phi_l ).  \ee

\begin{theorem}\label{thm:Backlundtransformation:pma-2dToda}
If $\tau_n=\tau_n(x,s)$ solves the bilinear 2D Toda lattice equation
\eqref{eq:b2dToda:pma-2dToda}, and $\sigma _n=\sigma _n(x,s)$
satisfies \be  \frac {\partial ^2 \sigma _n }{\partial s
\partial x}\,\sigma _n = \frac {\partial \sigma _n }{\partial x}\frac {\partial \sigma _n }{\partial
s}\,,\ \, \sigma_{n+1}\sigma_{n-1}=\sigma_n^2,
\label{eq:condforBacklundtransformation:pma-2dToda} \ee then the
function $\tilde \tau _n $ defined by \be \tilde \tau _n = \tilde
\tau _n(x,s)= \sigma_n(\alpha x,\alpha ^{-1}s)  \tau _n(\alpha
x,\alpha ^{-1}s), \label{eq:Backlundtransformation:pma-2dToda} \ee
where $\alpha$ is a non-zero real constant,
 presents another solution to the
bilinear 2D Toda lattice equation \eqref{eq:b2dToda:pma-2dToda}.
\end{theorem}

\noindent {\it Proof:} Under the first condition in
\eqref{eq:condforBacklundtransformation:pma-2dToda}, a direct
computation tells that
\[ (\frac {\part ^2 \tilde \tau _{n}} {\part s\part x}
\tilde \tau _{n}- \frac {\part \tilde \tau _{n}} {\part x} \frac
{\part \tilde \tau _{n}} {\part s} )(x,s) = [\sigma_n ^2( \frac
{\part ^2 \tau _{n}} {\part s\part x} \tau _{n} - \frac {\part \tau
_{n}} {\part x} \frac {\part \tau _{n}} {\part s} )](\alpha x,\alpha
^{-1}s).
 \]
Thus, the second condition in
\eqref{eq:condforBacklundtransformation:pma-2dToda} ensures that
\[\ba {l}
\quad\D  (\frac {\part ^2 \tilde \tau _{n}} {\part s\part x} \tilde
\tau _{n}- \frac {\part \tilde \tau _{n}} {\part x} \frac {\part
\tilde \tau _{n}} {\part s}
 -\tilde \tau _{n+1}\tilde \tau _{n-1}+\tilde \tau
_n^2)(x,s) \vspace{2mm}\\
\D  = [\sigma_n^2 ( \frac {\part ^2 \tau _{n}} {\part s\part x} \tau
_{n} - \frac {\part  \tau _{n}} {\part x} \frac {\part \tau _{n}}
{\part s} - \tau _{n+1} \tau _{n-1}+ \tau _n^2)](\alpha x,\alpha
^{-1}s)=0. \ea
\] The theorem is proved. \hfill $\Box$

This theorem provides us with an auto-B\"acklund transformation of
the bilinear 2D Toda lattice equation \eqref{eq:b2dToda:pma-2dToda}.
Generally, it also generates new solutions to the nonlinear 2D Toda
lattice equations \eqref{eq:2dToda:pma-2dToda} and
\eqref{eq:2ndformof2dToda:pma-2dToda}
 from a given solution
to the bilinear 2D Toda lattice equation
\eqref{eq:b2dToda:pma-2dToda}, through the transformations given in
the introduction.
 However, the case of $\alpha =1$
doesn't lead to new solutions to the nonlinear 2D Toda lattice
equation \eqref{eq:2dToda:pma-2dToda}.

A particular selection of $\sigma _n$ in Theorem
\ref{thm:Backlundtransformation:pma-2dToda} engenders the following
corollary.

\begin{col}\label{eq:specialDabouxTransform:pma-2dToda}
Let $\tau_n=\tau_n(x,s)$ be a solution to the bilinear 2D Toda
lattice equation \eqref{eq:b2dToda:pma-2dToda} and $\alpha $ be a
non-zero real constant. If $a_n(x)$ and $b_n(s)$ satisfy \be
a_{n+1}(x)a_{n-1}(x)b_{n+1}(s)b_{n-1}(s)=(a_n(x))^2(b_n(s))^2,
\label{eq:specialcondforBacklundtransformation:pma-2dToda} \ee then
$\tilde \tau_n$ with $\sigma_n(x,s)=a_n(x)b_n(s) $:
 \be \tilde \tau _n = \tilde \tau
_n(x,s)= a_n(\alpha x)b_n(\alpha ^{-1} s)  \tau _n(\alpha x,\alpha
^{-1}s) \label{eq:specialBacklundtransformation:pma-2dToda} \ee
solves the bilinear 2D Toda lattice equation
\eqref{eq:b2dToda:pma-2dToda}.

In particular, if $a(x)$, $b(s)$, $f(x)$ and $g(s)$ are real
functions but $f(x)$ and $g(s)$ are positive or negative, then
$\tilde \tau_n$ with $ a_n(x)=a(x)(f(x))^n$ and $
b_n(s)=b(s)(g(s))^n$: \be \tilde \tau_n = \tilde \tau _n(x,s)=
a(\alpha x)b(\alpha ^{-1}s)(f(\alpha x))^n(g(\alpha ^{-1}s))^n\tau
_n(\alpha x,\alpha ^{-1}s) \label{eq:specialtildetau:pma-2dToda} \ee
solves the bilinear 2D Toda lattice equation
\eqref{eq:b2dToda:pma-2dToda}.
\end{col}

In the above corollary, the assumption that $f(x)$ and $g(s)$ are
positive or negative is just to guarantee that $\tilde \tau _n$ is
well defined over the domain of $x,s\in \R $ and $n\in \Z$.

A combination of Theorems \ref{thm:gsoltau:pma-2dToda},
\ref{thm:CasoratianSolutions:pma-2dToda} and
\ref{thm:Backlundtransformation:pma-2dToda} offers us an approach
for constructing Casoratian type solutions to the bilinear 2D Toda
lattice equation \eqref{eq:b2dToda:pma-2dToda}.

If we take $a(x)=b(s)=1$, $f(x)=x^\beta $ and $g(s)=s^\alpha $, the
resulting solution $\tilde \tau _n$ with $\alpha =1$ gives the
solution presented in \cite{Hirota-PTPS1988}.

Let us take
 \be
\bar \phi_i=p_i^n\textrm{e}^{ \lambda _i x+\mu _is
+q_i}=\textrm{e}^{ -\varepsilon p_i^\delta x+ \varepsilon
p_i^{-\delta } s}  \phi_i ,\ 1\le i\le N, \ee where $p_i\ne 0$,
$q_i,$ $\lambda _i$ and $\beta _i$, $1\le i\le N$, are arbitrary
real constants. The set of functions $\{\phi _i \}_{i=1}^N$
satisfies \eqref{eq:cond1:pma-2dToda} and
\eqref{eq:cond2:pma-2dToda} with $A=\textrm{diag}(\lambda
_1,\cdots,\lambda _N)$ and $B=\textrm{diag}(\mu _1,\cdots,\mu _N)$
as showed before, and obviously, we have \begin{eqnarray} && \bar
\tau
_n=\textrm{Cas}(\bar \phi_1,\cdots, \bar \phi_N) \nonumber \\
&& =\exp \bigl(\,\sum_{i=1}^N (\lambda _ix +\mu _i s+q_i)\,\bigr)
\prod _{i=1}^N p_i^n \prod _{i>j}(p_i-p_j) \nonumber \\
&&  = \exp (-\varepsilon \sum_{i=1}^N p_i^\delta x )\exp(\varepsilon
\sum_{i=1}^N p_i^{-\delta} s) \textrm{Cas}(\phi_1,\cdots, \phi_N).
\label{eq:bartau:pma-2dToda} \end{eqnarray} The last equality in
\eqref{eq:bartau:pma-2dToda} also tells us a formula for $\tau _n
=\textrm{Cas}(\phi_1,\cdots,\phi_N)$, where the $\phi_i$'s are
defined by \eqref{eq:specialcase1ofphi_i:pma-2dToda} with $M=1$. It
follows from the above corollary with $\alpha =1$ that $\bar \tau
_n$ is a Casoratian solution to the bilinear 2D Toda lattice
equation \eqref{eq:b2dToda:pma-2dToda}. Again from the above
corollary, we have a class of Casoratian type solutions to the
bilinear 2D Toda lattice equation \eqref{eq:b2dToda:pma-2dToda}:
 \be \tilde \tau _n=  a(\alpha x)b(\alpha
^{-1}s)(f(\alpha x))^n(g(\alpha ^{-1}s))^n \exp \bigl(\,\sum_{i=1}^N
(\lambda _i\alpha x +\mu _i \alpha ^{-1} s+q_i)\,\bigr) \prod
_{i=1}^N p_i^n \prod _{i>j}(p_i-p_j) . \ee Obviously, these
solutions $\tilde \tau _n$ are all just special cases of
\eqref{eq:specialtildetau:pma-2dToda} with $\tau_n=1$. They generate
non-constant solutions to the nonlinear 2D Toda lattice equation
\eqref{eq:2ndformof2dToda:pma-2dToda}, but only the zero solution to
the nonlinear 2D Toda lattice equation \eqref{eq:2dToda:pma-2dToda}.

If we now take the functions $\phi_i$, $1\le i\le N$, in
\eqref{eq:specialcase1ofphi_i:pma-2dToda} with $M=2$, i.e., \be
\phi_i= p_{i1}^n\textrm{e}^{\,(\varepsilon p_{i1}^\delta +\lambda
_i) x-(\varepsilon p_{i1}^{-\delta }-\mu _i)s+q
_{i1}}+p_{i2}^n\textrm{e}^{\,(\varepsilon p_{i2}^\delta +\lambda _i)
x-(\varepsilon p_{i2}^{-\delta }-\mu _i)s+q _{i2}}, \ 1\le i\le N,
\ee then by the above corollary, we have a class of Casoratian type
solutions to the bilinear 2D Toda lattice equation
\eqref{eq:b2dToda:pma-2dToda}: \be \tilde \tau _n= \tilde \tau
_n(x,s)= a(\alpha x)b(\alpha ^{-1}s)(f(\alpha x))^n(g(\alpha
^{-1}s))^n\textrm{Cas}(\phi_1,\cdots,\phi_N)(\alpha x,\alpha
^{-1}s).
 \ee
 A general case of $M$ in
\eqref{eq:specialcase1ofphi_i:pma-2dToda} can produce more general
Casoratian type solutions to the bilinear 2D Toda lattice equation
\eqref{eq:b2dToda:pma-2dToda}. Such solutions $\tilde \tau _n$ can
also generate new solutions to the nonlinear 2D Toda lattice
equation \eqref{eq:2ndformof2dToda:pma-2dToda}, and if $\alpha \ne
1$, new solutions to the nonlinear 2D Toda lattice equation
\eqref{eq:2dToda:pma-2dToda}.

\section{\bf Concluding remarks}
\label{sec:Concludingremarks:pma-2dToda}

A general Casoratian formulation of the bilinear 2D Toda lattice
equation \eqref{eq:b2dToda:pma-2dToda}
 has been presented by means of
the bilinear form of \eqref{eq:b2dToda:pma-2dToda}. The resulting
theory provides us with an effective approach for constructing exact
solutions to the bilinear 2D Toda lattice equation
\eqref{eq:b2dToda:pma-2dToda}. Special classes of functions
satisfying \eqref{eq:cond1:pma-2dToda} and
\eqref{eq:cond2:pma-2dToda}, e.g., the functions defined by
\eqref{eq:specialcase1ofphi_i:pma-2dToda} and
\eqref{eq:specialcase2ofphi_i:pma-2dToda}, were used to generate
Casoratian solutions, and further using the B\"acklund
transformation in Theorem
\ref{thm:Backlundtransformation:pma-2dToda}, various examples of
Casoratian type solutions were presented.

We remark that the solutions $\tilde \tau _n$ presented in Corollary
\ref{eq:specialDabouxTransform:pma-2dToda} may not be exactly
Casoratian, even if $\tau _n$ is Casoratian. For example, $\tilde
\tau _n$ is non-Casoratian when $f(x)$ and $g(s)$ are not constant
functions. On the other hand, taking different types of
 functions for $a(x),$ $b(s),$ $f(x)$ and $g(s)$ can yield
 positon and complexiton type solutions.

There are also two other questions that we are interested in. The
first question is how to solve the system of differential-difference
equations in \eqref{eq:cond1:pma-2dToda} and
\eqref{eq:cond2:pma-2dToda} generally, in particular, in the case
where the conditions
\eqref{eq:condition2forspecialsolutionofphi_i:pma-2dToda} and
\eqref{eq:condition3forspecialsolutionofphi_i:pma-2dToda} are not
satisfied, or more generally, the equations
\eqref{eq:fundamentalsolutiononx:pma-2dToda} and
\eqref{eq:fundamentalsolutionons:pma-2dToda} don't hold. This will
bring us very different Casoratian solutions to the bilinear 2D Toda
lattice equation \eqref{eq:b2dToda:pma-2dToda}. The second question
is what kind of Casoratian formulations can exist for
Pfaffianization of discrete soliton equations
\cite{HuZT-JMAA2004,ZhaoLH-JPSJ2004}, for example, for
Pfaffianization of the 2D Toda lattice equation
\cite{HuZT-JMAA2004}. Any answers to these two questions will
enhance our understanding of both diversity of Casoratian type
solutions and university of Casoratian formulations.

\vskip 2mm

\small \noindent{\bf Acknowledgements:}
The work was in part supported by State Key
   Laboratory of Scientific and Engineering Computing, Chinese Academy of
   Sciences, Beijing, PR China and the University of South Florida, Tampa, Florida, USA.
I would also like to thank Professor Xing-Biao Hu for his warm
hospitality and invaluable discussions during my visit.

\small

\setlength{\baselineskip}{13pt}

\end{document}